# Development of a new generation of 3D pixel sensors for HL-LHC

G.-F. Dalla Betta[a,b,*], M. Boscardin[c,b], G. Darbo[d], R. Mendicino[a,b], M. Meschini[e], A. Messineo[f,g], S. Ronchin[c], D M S Sultan[a,b], N. Zorzi[c,b]

[a] *Università di Trento, Dipartimento di Ingegneria Industriale, I-38123 Trento, Italy*
[b] *TIFPA INFN, I-38123 Trento, Italy*
[c] *Fondazione Bruno Kessler, I-38123 Trento, Italy*
[d] *INFN Sezione di Genova, I-16146 Genova, Italy*
[e] *INFN Sezione di Firenze, I-50019 Sesto Fiorentino, Italy*
[f] *Università di Pisa, Dipartimento di Fisica, I-56127 Pisa, Italy*
[g] *INFN, Sezione di Pisa, I-56127 Pisa, Italy*

**Abstract**

This paper covers the main technological and design aspects relevant to the development of a new generation of thin 3D pixel sensors with small pixel size aimed at the High-Luminosity LHC upgrades.

## 1. Introduction

After their successful application to the ATLAS Insertable B-Layer (IBL) [1], and owing to their intrinsic radiation hardness, 3D pixel sensors are considered a viable option for the "Phase 2" upgrades at the High-Luminosity LHC (HL-LHC), in particular for the innermost tracking layers of ATLAS and CMS, which will have to cope with extreme radiation fluences (up to $2\times10^{16}$ $n_{eq}$ cm$^{-2}$). To this purpose, we are developing a new generation of 3D pixels optimized for increased pixel granularity (25×100 or 50×50 μm$^2$ pixel size), reduced material budget and better geometrical efficiency. Compared to the double-sided 3D sensors produced at FBK for the ATLAS IBL [2], these requirements call for a modified (single-sided) technology allowing for downscaled sensor dimensions: thinner active layers (~100 μm), narrower electrodes (~5 μm), reduced inter-electrode spacing (~30 μm), and very slim (~100 μm) or active edges.

## 2. Process development

Due to mechanical yield issues, we propose a new 3D structure made with a single-sided approach on Silicon-Silicon Direct Wafer Bonded (SiSi DWB) substrates from IceMOS Technology Ltd. (Belfast, UK), consisting of a p$^-$ Float Zone High-Resistivity (HR) layer of the desired thickness (e.g., 100 or 130 μm for sensors of the first batch) directly bonded to a p$^{++}$ Low-Resistivity (LR) handle wafer (see Fig. 1a).

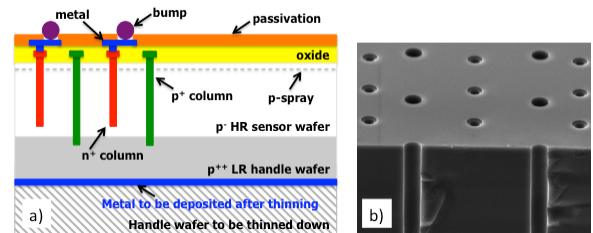

Figure 1 (a) Schematic cross-section of the proposed thin 3D sensors on SiSi DWB substrate, and (b) SEM micrograph of two sets of columns etched by DRIE (misalignment is intentional).

---

[*] Corresponding author. Tel.: +39-0461-283904; fax: +39-0461-281977; e-mail: gianfranco.dallabetta@unitn.it.

Besides providing a high mechanical stability, the LR handle wafer also allows the sensor bias to be applied from the back side, thus easing the front side layout. In fact, the $p^+$ columns are etched through the HR layer and penetrate the LR wafer, thus making a good ohmic contact that can be further improved by thinning the handle wafer and depositing a metal layer. The latter steps can be performed as a post-processing combined to the bump bonding process.

The $n^+$ (read-out) columns, isolated at the surface by a p-spray layer, are not etched completely through the HR layer, but they rather stop at a short distance (~15 μm) from the handle wafer in order to ensure a high breakdown voltage (higher than 100 V before irradiation), as already proved in existing devices [3] and confirmed by TCAD simulations for new ones.

Both types of columns are etched from the same wafer side (that was not the case for the previous double-sided process [3]). This modified approach was successfully proved at FBK (see Fig. 1b): a first set of narrow columns was etched by Deep Reactive Ion Etching (DRIE), followed by column partial filling with poly-Si. Then, a second set of wider columns was etched by DRIE without any problem.

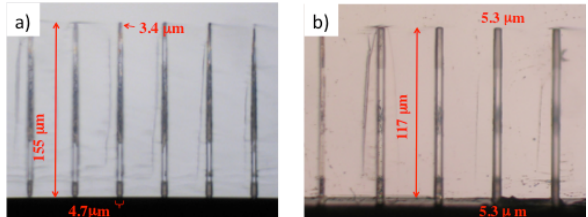

Figure 2 SEM micrographs of (a) ohmic columns, and (b) junction columns etched by DRIE.

The different depths of the two types of columns require different etching recipes. As an example, Fig. 2 shows the SEM micrograph of ohmic and junction columns aimed at the 130 μm HR layer thickness. The etching of ohmic columns (Fig. 2a) was optimized for depth in order to reach the LR handle wafer. This comes at the expense of a non-uniform column width that, however, is not critical for ohmic columns. On the contrary, width uniformity is essential for read-out columns to obtain a uniform electric field distribution. As can be seen from Fig. 2b, the final result is remarkably good. Similar tests with shorter columns were successfully performed also for the 100 μm HR layer thickness.

As for the pixel design, two different sizes are considered (see Fig. 3): 50×50 μm² with one $n^+$ column, and 25×100 μm², with two $n^+$ columns. The corresponding inter-electrode spacings (L) are ~35 μm and ~28 μm, respectively, making the 25×100 μm² pixel more radiation tolerant (a signal efficiency higher than 50% after $2\times10^{16}$ $n_{eq}$ cm$^{-2}$ has been estimated by TCAD simulations and by projections based on existing data [4]). However, the 25×100 μm² pixel, due to the presence of two read-out columns, exhibits a larger capacitance (~100 fF, to be compared to ~50 fF for the 50×50 μm² pixel) with impact on the noise. Moreover this layout is more critical, since the bump-bonding pad must be placed very near to both $n^+$ and $p^+$ columns (Fig. 3b). Alternative designs, featuring bonding pads on top of the columns, will also be tested in collaboration with bump bonding facilities (Selex and IZM).

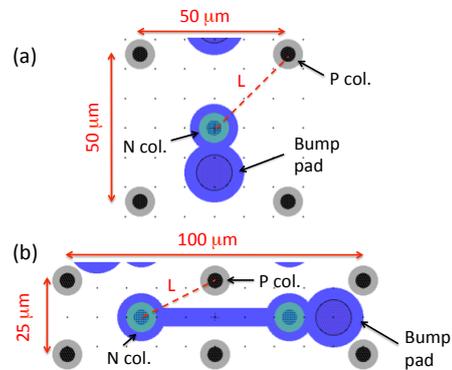

Figure 3 Layouts of (a) 50×50 μm² and (b) 25×100 μm² 3D pixels.

## 3. Conclusion

We have reported on the key process steps enabling the development of a new generation of 3D pixel sensors with small pixel size and thin active layers. Fabrication of a first batch of these detectors is under way at FBK.

### Acknowledgment

This work was supported by the Italian National Institute for Nuclear Physics (INFN), Projects ATLAS, CMS, RD-FASE2 (CSN1), and by AIDA-2020 project EU-INFRA proposal no. 654168.